\documentclass[]{article}
\usepackage{geometry}                % See geometry.pdf to learn the layout options. There are lots.
\usepackage{graphicx}
\usepackage{amssymb}
\usepackage{epstopdf}
\usepackage{subfigure}
\usepackage{url}

\begin{document}

\title{Generative complexity of Gray-Scott model}
\author{Andrew Adamatzky\\University of the West of England, Bristol, United Kingdom}

\maketitle

\begin{abstract}
In the Gray-Scott reaction-diffusion system one reactant is constantly fed in the system, another reactant is reproduced by consuming the supplied reactant and also converted to an inert product.
The rate of feeding one reactant in the system and the rate of removing another reactant from the system determine configurations of concentration profiles: stripes, spots, waves. We calculate the generative 
 complexity --- a morphological complexity of concentration profiles grown from a point-wise perturbation of the medium --- of the Gray-Scott system for a range of the feeding and removal rates. 
 The morphological complexity is evaluated using  Shannon entropy,  Simpson diversity, approximation of Lempel-Ziv complexity, and expressivity (Shannon entropy divided by space-filling). 
 We analyse behaviour of the systems with highest values of the generative morphological complexity and show that the Gray-Scott systems expressing highest levels of the complexity 
 are composed of the  wave-fragments (similar to wave-fragments in sub-excitable media) and travelling localisations (similar to quasi-dissipative solitons and gliders in Conway's Game of Life). 
%\emph{Keywords:} reaction-diffusion model, Gray-Scott model, complexity, Shannon entropy, expressivity 
 \end{abstract}

%\section{}
%\subsection{}

\section{Introduction}

The Gray-Scott model~\cite{gray1983autocatalytic, lee1993pattern, pearson1993complex} is a system of two reactants $U$ and $V$: the reactant $U$ is fed into the system, the reactants $V$  is present in the system initially,
 one molecule of $U$ reacts with  two molecules of $V$ producing three molecules of $V$. The model bears a striking similarity to the Lotka-Volterra model~\cite{lotka1920analytical}, where $U$ is a prey, $V$ is a predator and the Sel'kov model of glycolisis~\cite{sel1968self}, where $U$ is a substrate, $V$ is a product; analogy with two-variable Oregonator model of Belousov-Zhabotinsky medium~\cite{noyes1984alternative, jahnke1989chemical}, 
  where $U$ is a catalist and $V$ is activator, are less obvious  however spatio-temporal dynamics  is often matching. The spatially extended Gray-Scott model with low coefficients of reactants diffusion shows a  rich variety of concentration profile patterns: strips, spots, waves~\cite{lee1993pattern, pearson1993complex}.  Concentration patterns which attracted most attention 
   include spots and auto-solitons~\cite{wei2001pattern, muratov2002stability, chen2011stability, hayase1997collision, nishiura2005scattering}, rings~\cite{morgan2004axisymmetric}, self-replicating patterns~\cite{lee1997replicating, nishiura2000self, hayase2000self, munteanu2006pattern, reynolds1997self},    stripes~\cite{lee1995lamellar, kolokolnikov2006zigzag},  spiral waves~\cite{farr1992rotating}. 
   The patterns are governed by a rate of feeding $U$ and a rate of removal of $V$. Pearson~\cite{pearson1993complex} proposed a 
   phenomenological classification of Gray-Scott model based of configurations of concentration profiles. The Pearson  classification was detailed and extended by Munafo~\cite{munafo2014stable, munafo2016} and 
   mapping between the Pearson-Munafo classes and Wolfram's classes of elementary cellular automata~\cite{wolfram1983statistical} has been proposed.   Many interesting results have been obtained with Gray-Scott model but no evaluation of its complexity has been done so far. We decided to fill the gap and analyse a generative morphological complexity of the system. The morphological complexity is evaluated via diversity of the configurations of concentration profiles using Shannon entropy, Simpson diversity, Lempel-Ziv complexity. To avoid parameterisation of initial random conditions we considered only the generative complexity -- the diversity of patterns 
 developed from a point-wise local perturbation of otherwise resting medium. This our approach is already proved to be efficient in studying complexity of cellular automata, and
  discrete models of excitable systems and populations~\cite{adamatzky2010generative, adamatzky2012phenomenology, adamatzky2010minimal}. 
   
\section{Gray-Scott model}

The Gray-Scott model~\cite{pearson1993complex} is comprised of 
two reactants $U$ and $V$ reacting as follows:
\begin{eqnarray*}
{\rightarrow} U \\
U+2V \rightarrow 3V  \\
V \rightarrow P 
\end{eqnarray*}
where $P$ is inert product, reactant $U$ is fed with rate $k$, reactant $V$ is converted to inert product $P$  with rate $F$,  $U$ reacts with $V$ with rate 1. 
The corresponding reaction-diffusion equations for concentrations $u$ and $v$ are 
\begin{eqnarray*}
\frac{\partial u}{\partial t} = D_u \nabla^2 u - uv^2 +F(1-u)\\
\frac{\partial v}{\partial t} = D_v \nabla^2 v + uv^2 - (F+k) v
\end{eqnarray*}

We integrated the system using forward Euler method with five-node Laplace operator, time step 1 and diffusion coefficients $D_u=2\times 10^{-5}$ and $D_v=10^{-5}$; 
these parameters have been chosen to stay compatible with \cite{pearson1993complex}. We evaluated  complexity measures by taking a grid of $256 \times 256$ nodes, each node $x$ but four assigned concentration values 
$u_x=1$ and $v_x=0$,  four neighbouring nodes at the centre of the lattice assigned $v_x=1$: 
$
\begin{array}{c|c}
1 & 1 \\ \hline
1 & 1 
\end{array}
$ 

For a given pair $(k,F)$ the grid allowed to evolve until propagation of the perturbation, 
measured as  $v>0.3$, reached a boundary of the grid, or no changes between two subsequent concentration profiles observed, or a number of iterations exceeded $10^3$. 
The measures were calculated on concentration profiles after the halting.

\section{Complexity measures}

\begin{figure}[!tbp]
\centering
\subfigure[]{\includegraphics[scale=0.33]{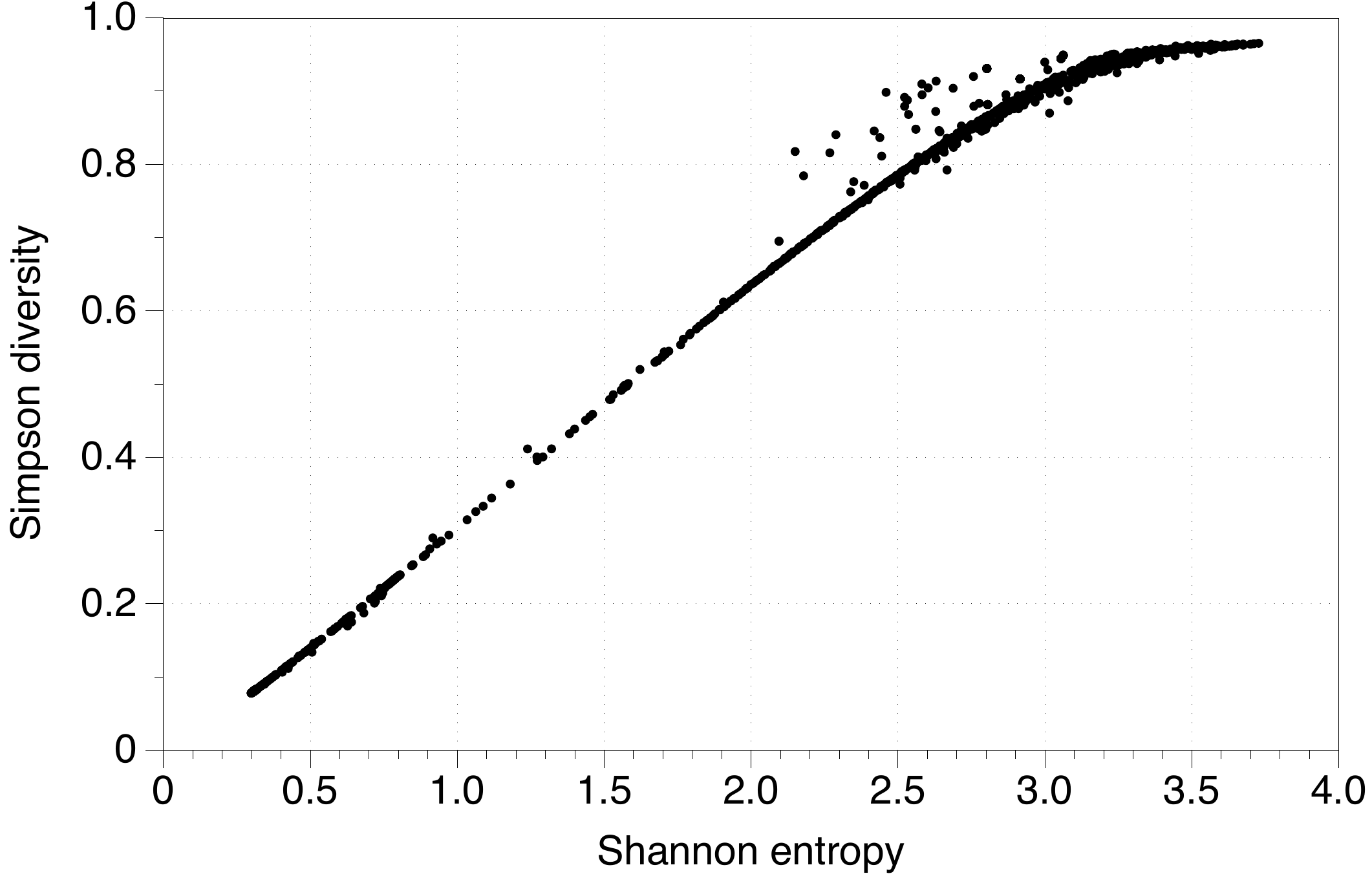}}
\subfigure[]{\includegraphics[scale=0.33]{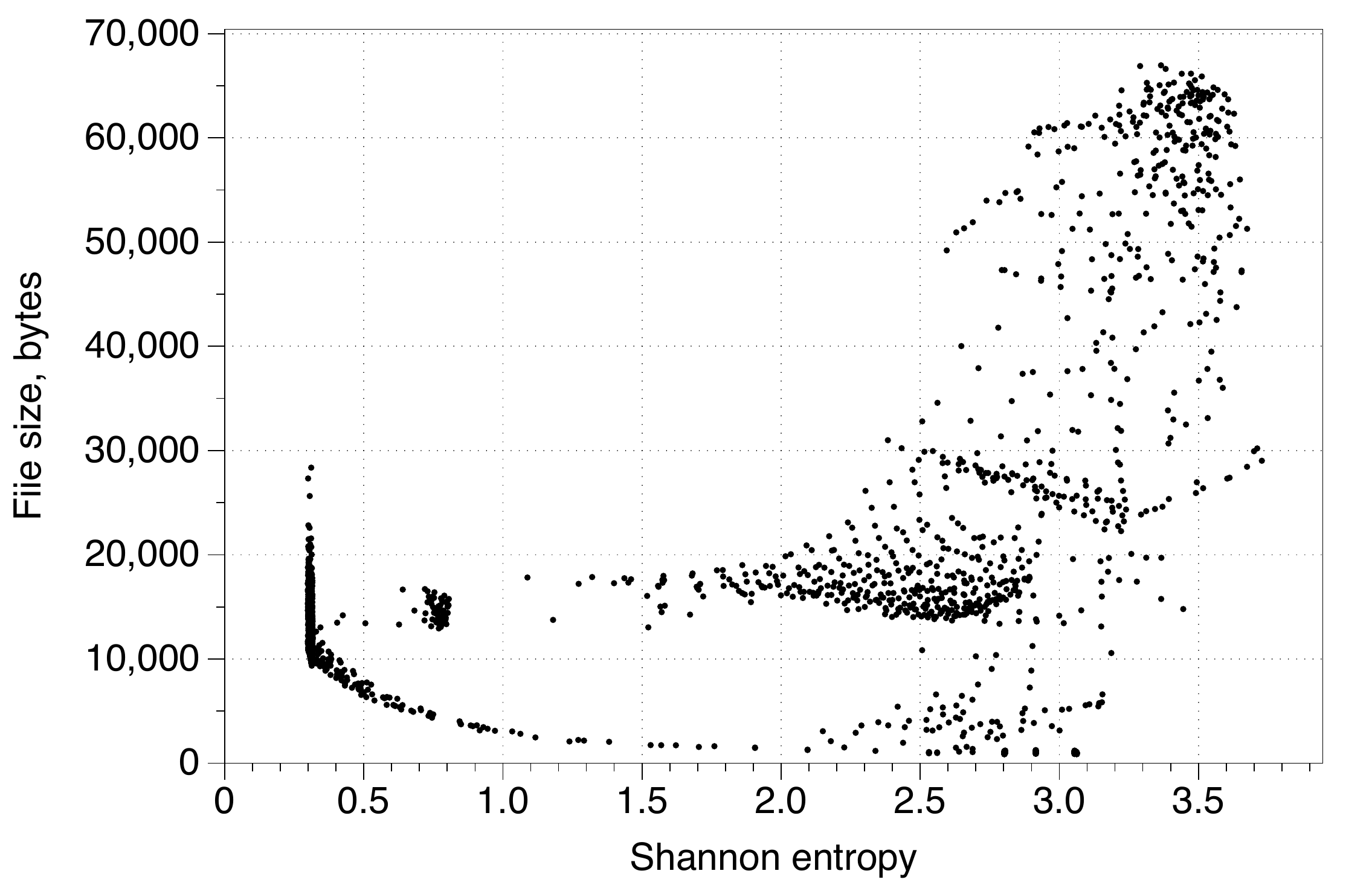}}
\subfigure[]{\includegraphics[scale=0.33]{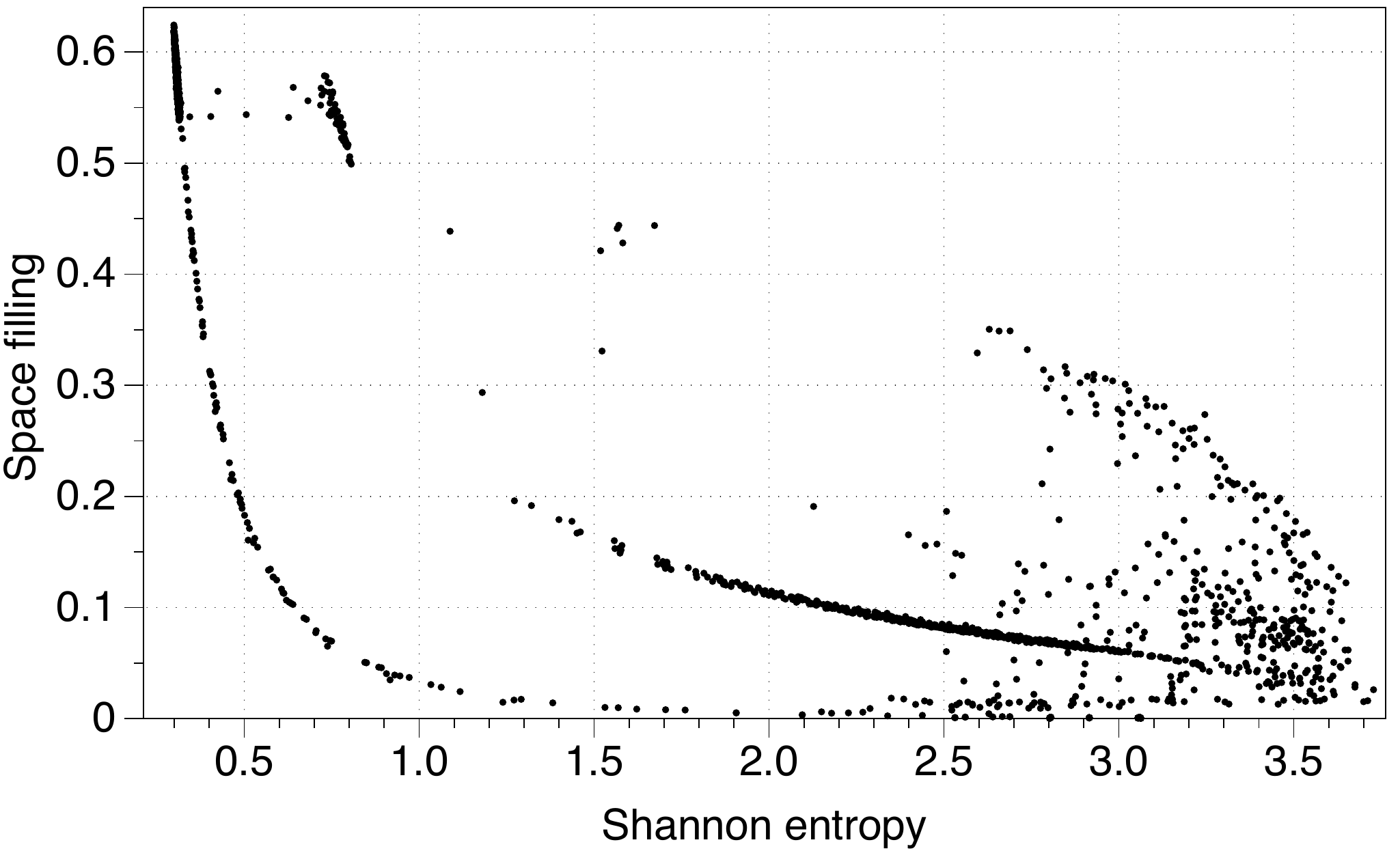}}
\caption{Shannon entropy $H$ versus 
(a) Simpson diversity $S$,  
(b) $LZ$ complexity,
(c ) Space filling $D$.}
\label{shannongraphs}
\end{figure}

We evolved the systems and evaluated complexities for 8320 pairs $(k,F)$, where  $k \in [0.020,0.072]$, $F \in [0.010,0.17]$, increments $0.001$. 

When evaluating complexity measures we binarized concentration profile of $V$ as follows. The  $256 \times 256$ nodes grid of concentrations is mapped onto an array $L$ of 
$256 \times 256$ cells,  where each cell $x$ is assigned value `1' if the concentration of $V$ at the corresponding grid node $x$ exceeds 0.3; otherwise the cell is assigned value `0'. 
 Let  $W=\{ 0,1 \}^9$ be a set of all possible configurations of a 9-node neighbourhood $B_x$ including the central node $x$. 
 Let $B$ be a configuration of matrix $L$, we calculate a number of non-quiescent  neighbourhood 
configurations as $\eta = \sum_{ x \in L} \epsilon(x)$, 
where $\epsilon(x)=0$ if for every resting $x$ all 
its neighbours are resting, and $\epsilon(x)=1$ otherwise. 

The Shannon entropy $H$ is calculated as 
$H =- \sum_{w \in W} (\nu(w)/\eta \cdot ln (\nu(w)/\eta))$, 
where $\nu(w)$ is a number of times the neighbourhood 
configuration $w$ is found in configuration $B$. 

Simpson's  diversity $S$ is calculated as $S=\sum_{w \in W} (\nu(w)/\eta)^2$. 
Simpson diversity linearly correlates with Shannon entropy for $H<3$; relationships becomes logarithmic for higher values of $H$ (Fig.~\ref{shannongraphs}a).

Lempel-Ziv complexity (compressibility) $LZ$ is evaluated by a size of PNG files of the configurations, this is sufficient because the 'deflation' algorithm used in 
PNG lossless compression~\cite{roelofs1999png,howard1993design, deutsch1996zlib}  is a variation of the classical Lempel--Ziv 1977 algorithm~\cite{ziv1977universal}.
There is a weak correlation between $H$ and $LZ$ for $H>3$ (Fig.~\ref{shannongraphs}b) therefore we will be considering these measures independently.

Space filling $D$ is a ratio of non-zero entries in $B$ to the total number of cells/nodes. This is used to estimate expressiveness. 
$D$ decreases by the power low with increase of $H$ when $H<1.5$,  and linearly $1.5 \leq H \leq 3$; 
there is a weak correlation between $D$ and $H$ for high values of entropy, $H>3$ (Fig.~\ref{shannongraphs}c).

Expressiveness is calculated as the Shannon entropy $H$ divided by space-filling ratio $D$, the expressiveness reflects the `economy of diversity'.

\section{Rules with highest generative complexity}

\begin{figure}[!tbp]
\centering
\includegraphics[width=0.8\textwidth]{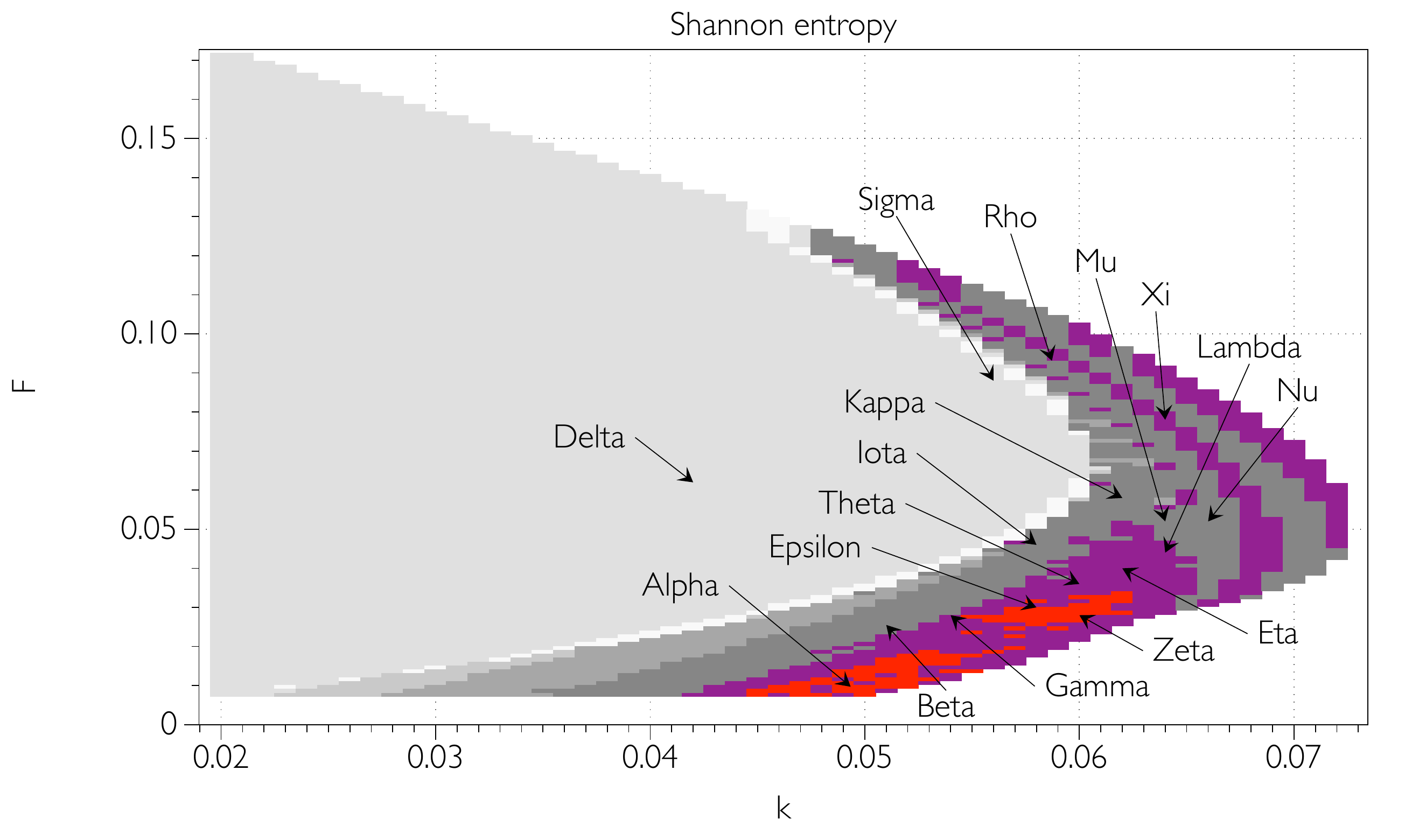}
\caption{Heat map of the Shannon entropy values for $(k,F)$ pairs.
Red indicates $H \in [3.5,4]$, magenta $H \in [3,3.5)$, the rest are gradations of grey indicated lower values of the entropy. 
Approximate positions of the Pearson-Munafo classes~\cite{munafo2016} are shown.indicated.} 
\label{munafoonshannon}
\end{figure}

\begin{table}[!tbp]
\centering
\caption{Values of $k$ and $F$ for top five measures of Shannon entropy $H$, 
Simpson diversity $S$, approximation of Lempel-Ziv complexity $LZ$, and expressivity $E$. }
\begin{subtable}
\centering
\begin{tabular}{c|cc}
$H$	&	$k$	&	$F$	\\ \hline
3.7278943	&	0.048	&	0.014	\\   
3.7112935	&	0.046	&	0.011	\\   
3.6995924	&	0.045	&	0.01	\\ 
3.6750243	&	0.048	&	0.013	\\   
3.6747382	&	0.047	&	0.013	\\  
\end{tabular}
\end{subtable}
\begin{subtable}
\centering
\begin{tabular}{c|cc}
$S$ 	&	$k$	&	$F$	\\ \hline
0.9652938	&	0.048	&	0.014	\\   
0.96467775	&	0.046	&	0.011	\\   
0.96432036	&	0.047	&	0.013	\\   
0.9642514	&	0.049	&	0.015	\\   
0.96407634	&	0.062	&	0.036	\\ 
\end{tabular}
\end{subtable}\\
\vspace{5mm}
\begin{subtable}
\centering
\begin{tabular}{c|cc}
   $LZ$	&	$k$	&	$F$	\\ \hline
66960	&	0.06	&	0.027	\\   
66889	&	0.06	&	0.026	\\   
66614	&	0.056	&	0.027	\\   
66160	&	0.055	&	0.023	\\   
66152	&	0.058	&	0.023	\\
\end{tabular}
\end{subtable}
\begin{subtable}
\centering
\begin{tabular}{c|cc}
$E$	&		$k$	&	$F$	\\ \hline
264.8449	&	0.049	&	0.01	\\   
251.12997	&	0.053	&	0.017	\\   
241.7313	&	0.045	&	0.01	\\   
229.02385	&	0.046	&	0.011	\\   
226.3538	&	0.05	&	0.01	\\  
\end{tabular}
\end{subtable}
\label{classes}
\end{table}%

\begin{figure}[!tbp]
\centering
\includegraphics[width=0.8\textwidth]{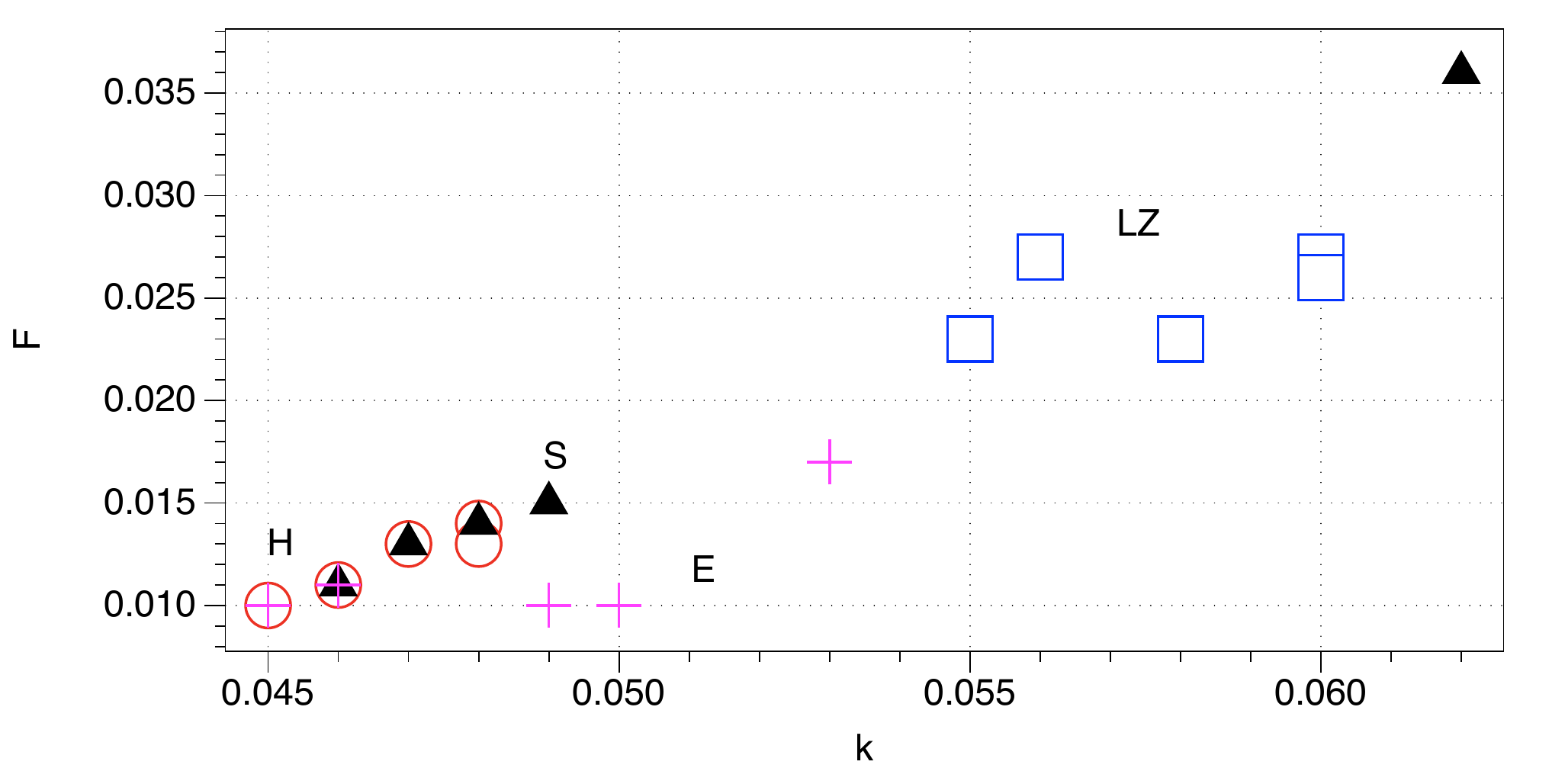}
\caption{
Top five pairs $(k, F)$ producing most complex concentration profiles for each measure of complexity, as in Tab.~\ref{classes}, 
are plotted in $k$-$F$ plane. Empty circles for Shannon entropy $H$, filled triangles for Simpson diversity $S$, empty squares for $LZ$, and crosses for expressivity $E$. 
}
\label{topfivepolot}
\end{figure}

 Top five pairs $(k,F)$ responsible for generating patterns with highest Shannon entropy $H$, Simpson diversity $S$, approximation of Lempel-Ziv $LZ$ complexity, 
 and expressivity $E$ are shown in Tab.~\ref{classes} and plotted $k$-$F$ plane in Fig.~\ref{topfivepolot}.  Pairs with highest $H$ and $S$ form a compact cluster 
 in the domain $[0.045,0.01]\times[0.049,0.015]$  with the exception of one pair for $S$ being at $(0.062,0.036)$. Rules with highest $LZ$ also group compactly in 
 the domain  $[0.055,0.023] \times [0.06,0.027]$. The pairs $(k, F)$  corresponding to highest expressivity $E$ are rather widely spread along $k$-axis, 
 from $k=0.045$ to $k=0.053$ with $0.007$ units elevation up in $F$-axis, from $F=0.01$ to $0.017$. 
 The pair $(k=0.046, F=0.011)$ shows highest values of three  complexity measures: $H$, $S$ and $E$.

\begin{figure}[!tbp]
\begin{center}
\subfigure[$k$=0.048, $F$=0.014, $t$=6970]{\includegraphics[scale=0.13]{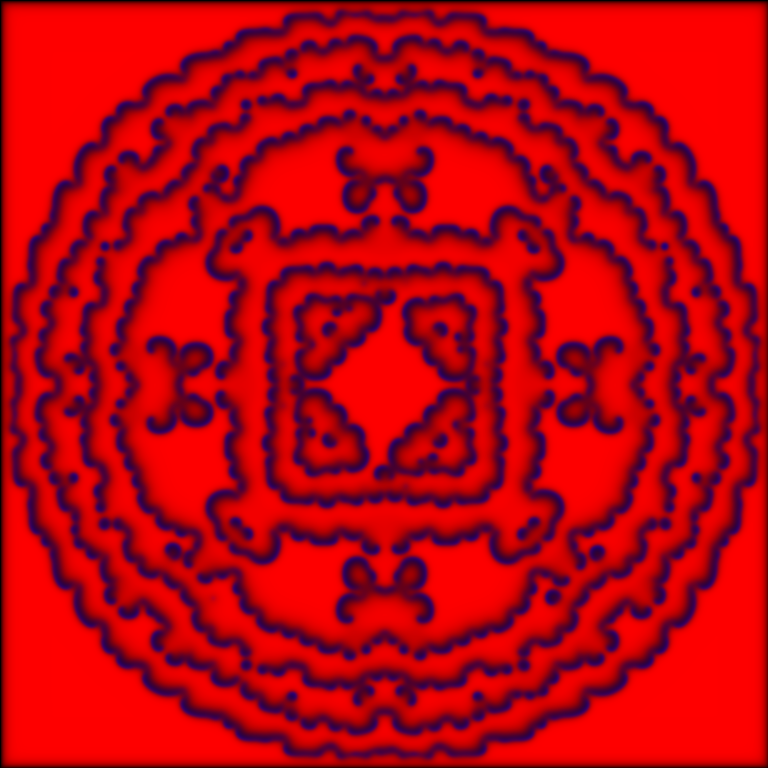}} 		%a  H, S
\subfigure[$k$=0.046, $F$=0.011, $t$=6040]{\includegraphics[scale=0.13]{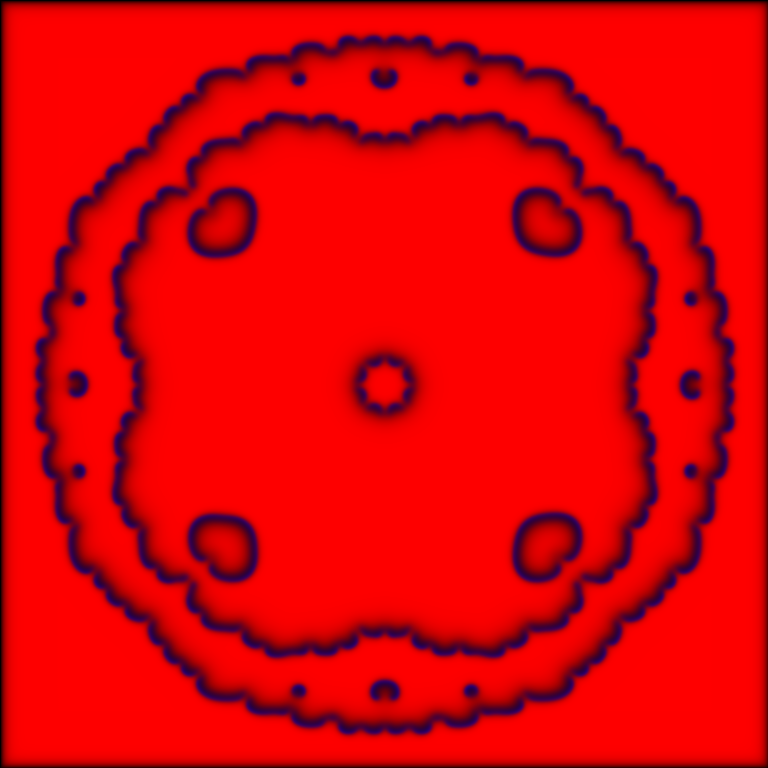}} 		%b  H, S, E  ---> most matches !!!!		%c H, E
\subfigure[$k$=0.048, $F$=0.013, $t$=5990]{\includegraphics[scale=0.13]{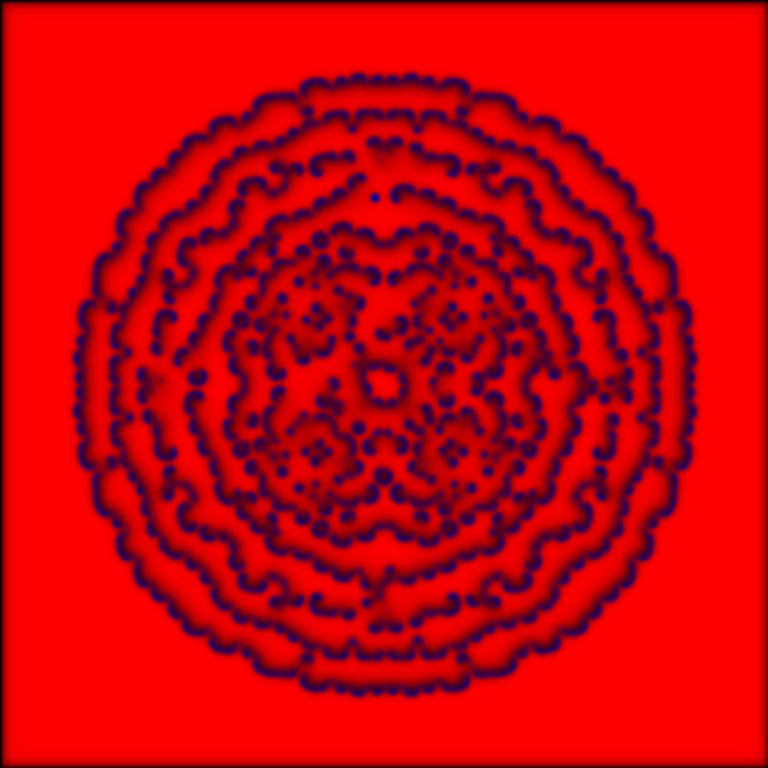}}  	%d H
\subfigure[$k$=0.049, $F$=0.015, $t$=6050]{\includegraphics[scale=0.13]{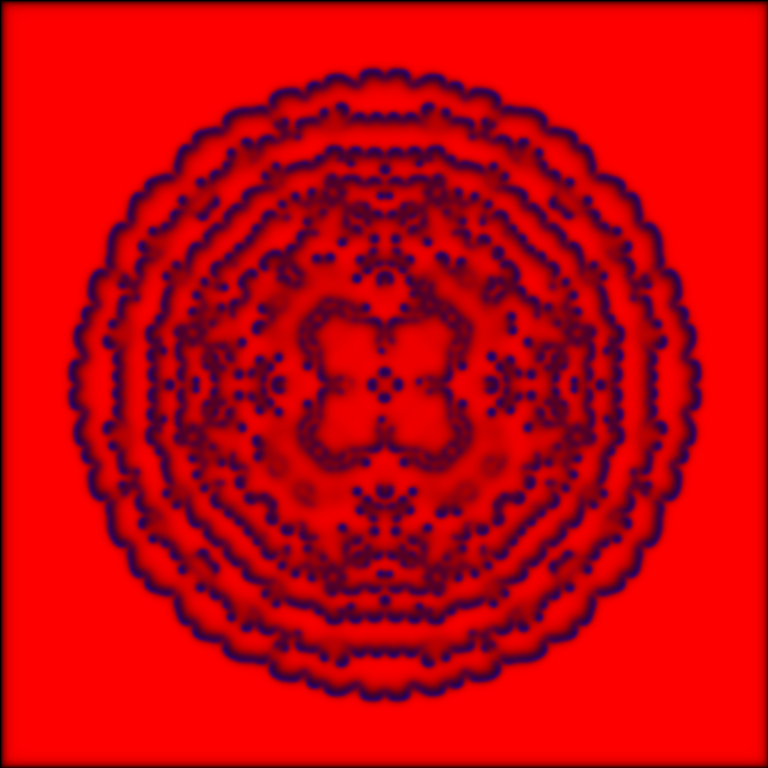}} 		%e S
\subfigure[$k$=0.060, $F$=0.027, $t$=8000]{\includegraphics[scale=0.13]{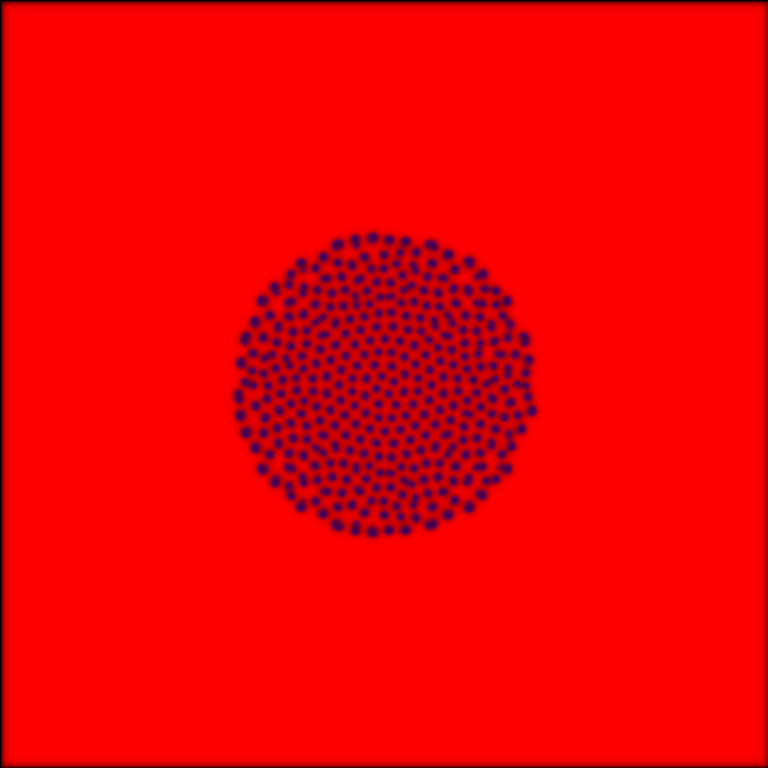}} 		%f LZ
\subfigure[$k$=0.060, $F$=0.026, $t$=8000]{\includegraphics[scale=0.13]{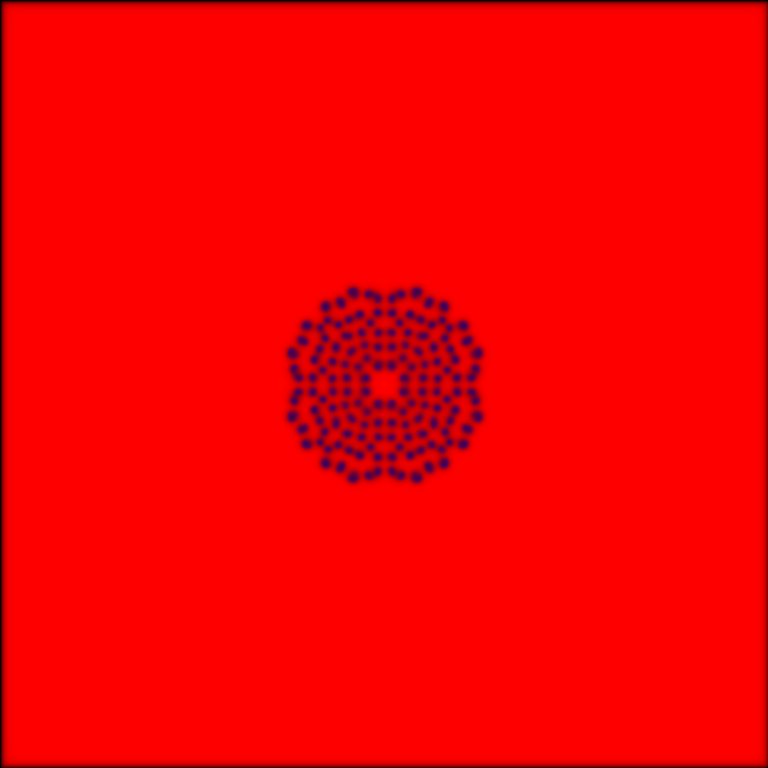}} 		%g LZ
\subfigure[$k$=0.056, $F$=0.027, $t$=8000]{\includegraphics[scale=0.13]{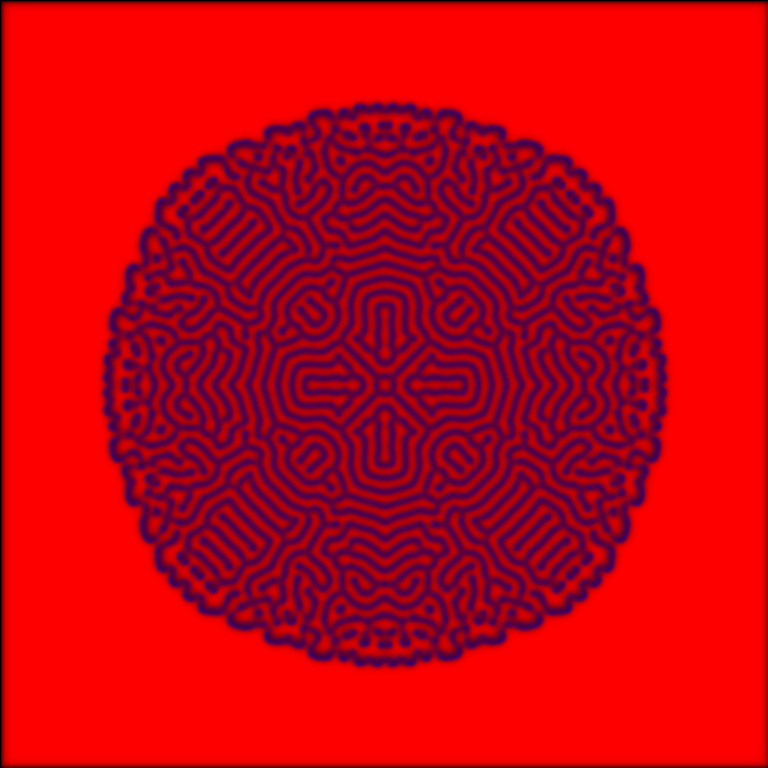}} 		%h LZ
\subfigure[$k$=0.055, $F$=0.023, $t$=8000]{\includegraphics[scale=0.13]{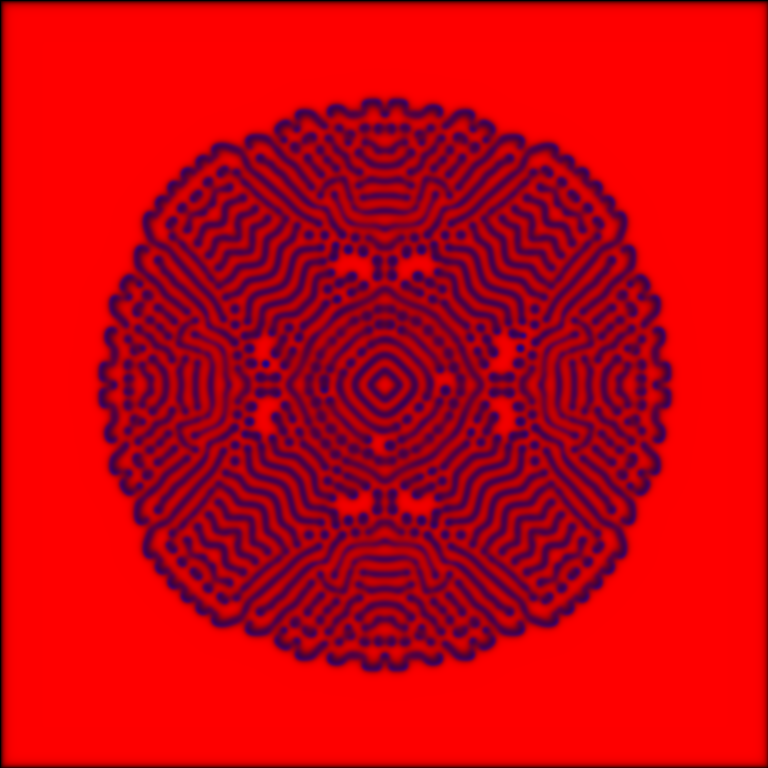}} 		%i LZ
\subfigure[$k$=0.058, $F$=0.023, $t$=8000]{\includegraphics[scale=0.13]{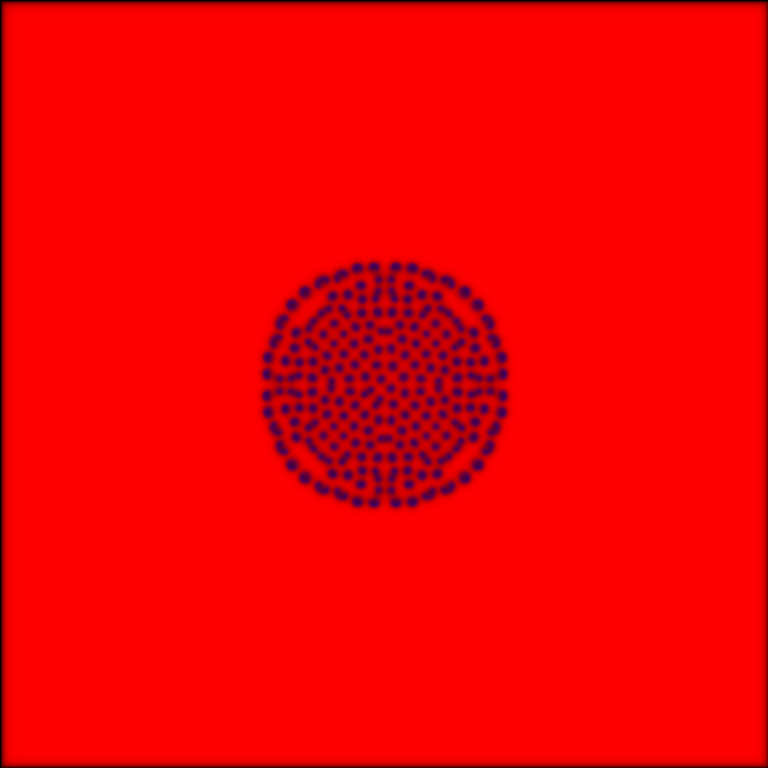}} 		%j LZ
\subfigure[$k$=0.049, $F$=0.010, $t$=8080]{\includegraphics[scale=0.13]{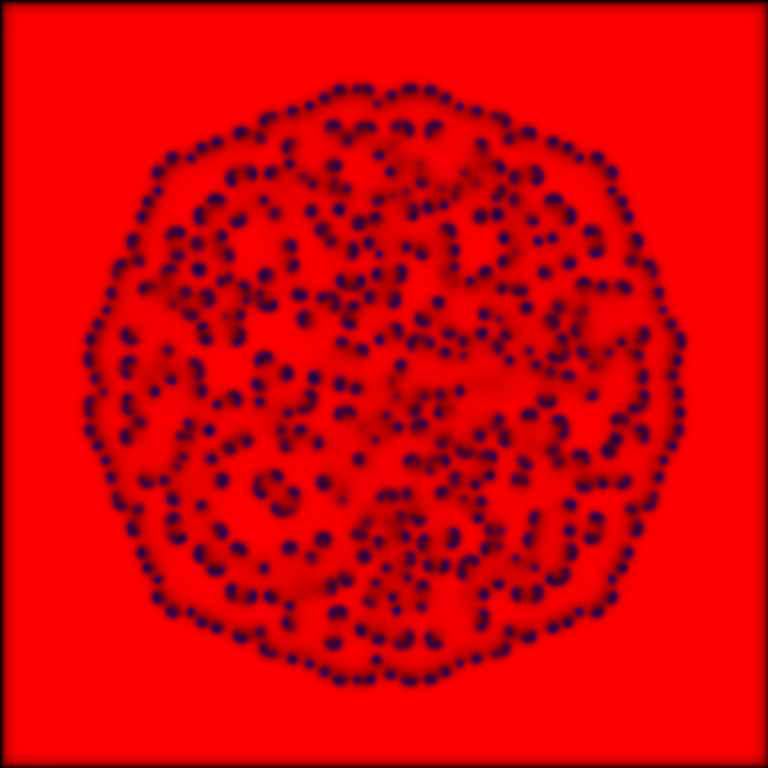}}  		%k E
\subfigure[$k$=0.053, $F$=0.017, $t$=8000]{\includegraphics[scale=0.13]{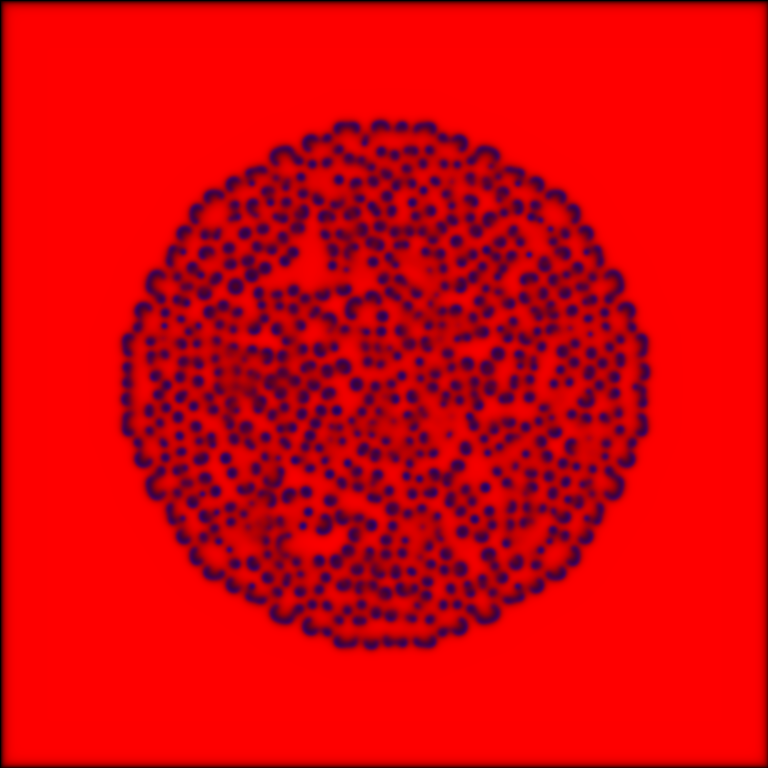}}  		%l E
\subfigure[$k$=0.050, $F$=0.010, $t$=8000]{\includegraphics[scale=0.13]{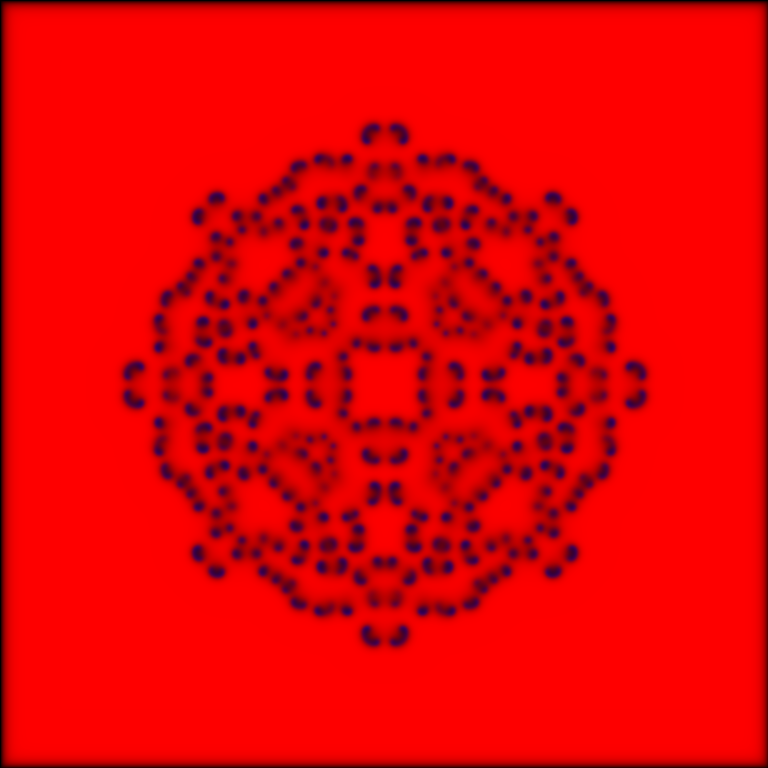}}  		%m E
\caption{Exemplar configurations for highest values of 
(a--e)  Shannon entropy $H$;
(a, b, e)  Simpson diversity $S$;
(f--j)  approximation $LZ$;
(k--m)  expressivity $E$.  Concentrations of $U$ and $V$ in each node $x$ are converted to RGB colour of the corresponding pixel $x$
as $(R, G, B)=(u_x \cdot 255, 0, v_x \cdot 255)$. Scale 0.1 of original size. See URLs to videos in Section ``Supplementary material". 
}
\label{examples}
\end{center}
\end{figure}

Exemplary snapshots of the concentrations profiles of pairs from Tab.~\ref{classes} are shown in Fig.~\ref{examples} and URLs to videos are listed in Section ``Supplementary material". 

\begin{figure}[!tbp]
\begin{center}
\subfigure[$t$=1680]{\includegraphics[scale=0.23]{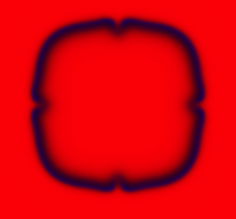}} %a
\subfigure[$t$=2340]{\includegraphics[scale=0.23]{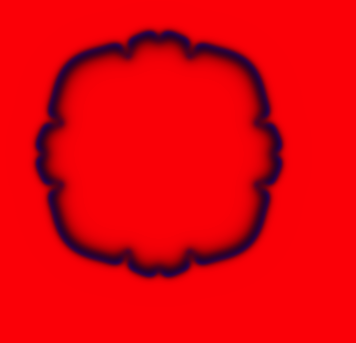}} %b
\subfigure[$t$=2740]{\includegraphics[scale=0.23]{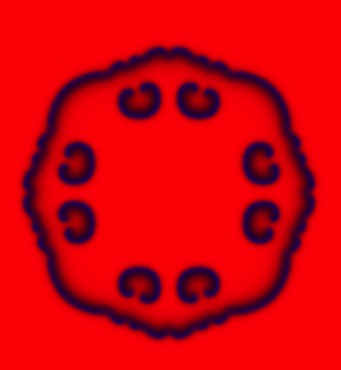}} %c
\subfigure[$t$=2990]{\includegraphics[scale=0.23]{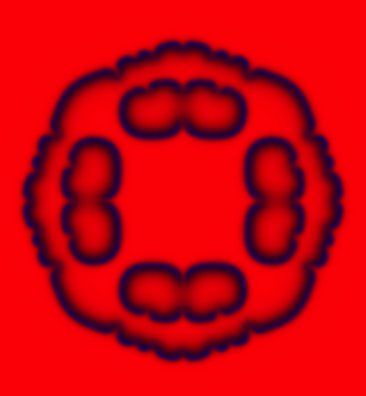}} %d
\subfigure[$t$=3430]{\includegraphics[scale=0.23]{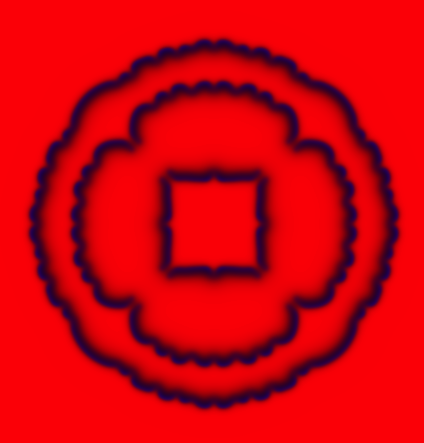}} %e
\subfigure[$t$=4100]{\includegraphics[scale=0.23]{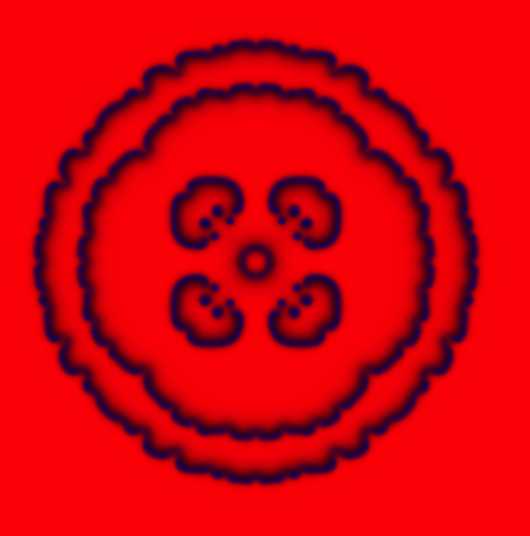}} %f
\subfigure[$t$=4300]{\includegraphics[scale=0.23]{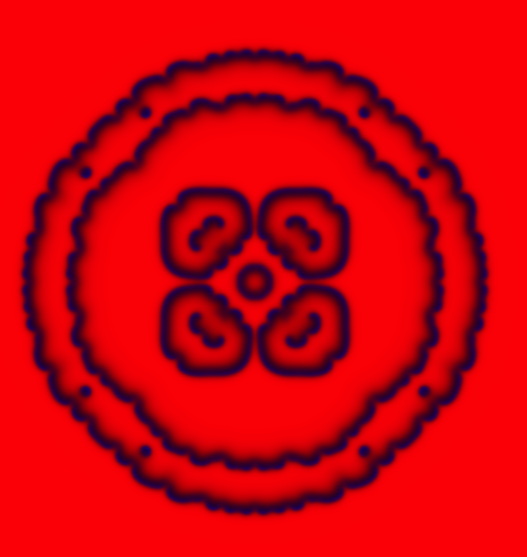}} %g
\subfigure[$t$=4920]{\includegraphics[scale=0.23]{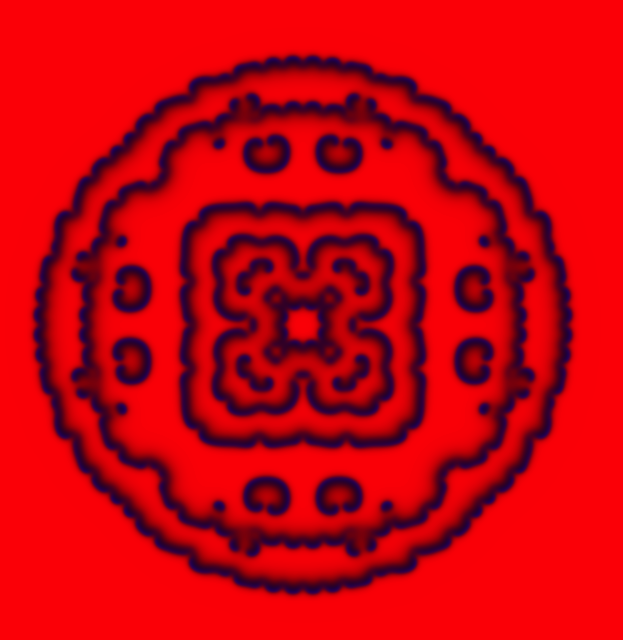}} %h
\caption{Snapshots of the medium's evolution governed by $(k, F)=(0.048, 0.014)$, $t$ is iteration at which the snapshot was recorded.  
Scale 0.23 of original size.}
\label{k_0_048_F_0_014}
\end{center}
\end{figure}

In the medium governed by $(k, F)$ pair with largest $H$ (Tab.~\ref{classes}, subtable $H$) the initial perturbation leads to formation of the 
circular wave-front propagating centrifugally (Fig.~\ref{k_0_048_F_0_014}a). 
After c. $2 \cdot 10^3$ iterations the wave-front loses its stability in four loci corresponding to centres of edges of the original perturbation (Fig.~\ref{k_0_048_F_0_014}b).
This causes four domains of the wave-front to propagate faster than the rest of the wave-front  (Fig.~\ref{k_0_048_F_0_014}c).
 Loci between the fast moving domains and the slow moving domains travelling centripetally form eight wave-fragments (Fig.~\ref{k_0_048_F_0_014}d). 
The eight wave-fragments fold into circular wave-fronts (Fig.~\ref{k_0_048_F_0_014}e) and then merge into two wave-fronts: one propagates away from t
he centre, another towards the centre. The centripetal wave-front collapses and produces four scroll wave-fragments travelling away from the centre (Fig.~\ref{k_0_048_F_0_014}f).
The scroll waves produce daughter scroll waves (Fig.~\ref{k_0_048_F_0_014}g).
Meantime centrifugal wave-front produces more centripetal wave-fragments  (Fig.~\ref{k_0_048_F_0_014}h). 
The process continues till the space is filled with interacting, annihilating and re-producing wave-fragments (Fig.~\ref{examples}a).

The medium governed by $(0.046, 0.011)$ produces patterns with 2nd highest $H$ and also included in the top five rules with highest $S$ and $E$. 
Behaviour of the medium is similar to that governed by $(0.048, 0.014)$ with minor variations (see video in Section ``Supplementary material"), e.g. it takes 
$\time 5$ more time for the initially formed circular wave-front to lose its stability and to start produce centripetal wave-fragments.

\begin{figure}[!tbp]
\begin{center}
\subfigure[$t$=1000]{\includegraphics[scale=0.5]{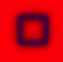}} %a
\subfigure[$t$=1500]{\includegraphics[scale=0.5]{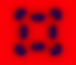}} %b
\subfigure[$t$=1790]{\includegraphics[scale=0.5]{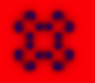}} %c
\subfigure[$t$=1990]{\includegraphics[scale=0.5]{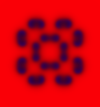}} %d
\subfigure[$t$=3210]{\includegraphics[scale=0.5]{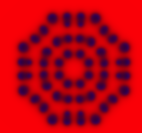}} %e
\subfigure[$t$=4000]{\includegraphics[scale=0.5]{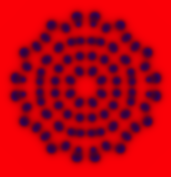}} %f
\subfigure[$t$=4860]{\includegraphics[scale=0.5]{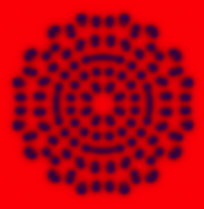}} %g
\subfigure[$t$=6000]{\includegraphics[scale=0.5]{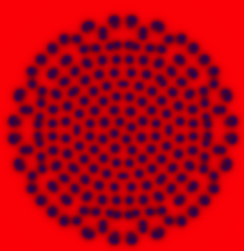}} %h
\caption{Snapshots of the evolution of the medium governed by $(k, F)=(0.060, 0.027)$, $t$ is iteration at which the snapshot was recorded. Scale 0.5 of original size.}
\label{k_0_060_F_0_027}
\end{center}
\end{figure}

The pair $(k, F)=(0.060, 0.027)$ produces concentration profiles with highest $LZ$ (Tab.~\ref{classes}, subtable $S$). 
Several snapshots of the medium evolution are shown in Fig.~\ref{k_0_060_F_0_027}. Initial perturbation gives rise to a circular wave-front (Fig.~\ref{k_0_060_F_0_027}a). 
The wave-front loses its stability after $1.5 \cdot 10^3$ iterations and produces four wave-fragments (Fig.~\ref{k_0_060_F_0_027}b). 
These wave-fragments become unstable and divide into two wave-fragments each (Fig.~\ref{k_0_060_F_0_027}c). 
Formation of the new wave-fragments and their multiplication continues (Fig.~\ref{k_0_060_F_0_027}c--h) till there is a space available.

\begin{figure}[!tbp]
\begin{center}
\subfigure[$t$=330]{\includegraphics[scale=0.5]{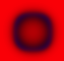}} %
\subfigure[$t$=540]{\includegraphics[scale=0.5]{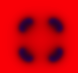}} %
\subfigure[$t$=840]{\includegraphics[scale=0.5]{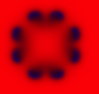}} %
\subfigure[$t$=1250]{\includegraphics[scale=0.5]{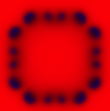}} %
%\subfigure[$t=$]{\includegraphics[scale=0.5]{figs/k_0_049_F_0_010/000001660.png}} %
\subfigure[$t$=2030]{\includegraphics[scale=0.5]{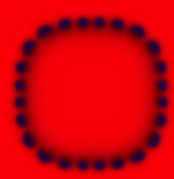}} %
\subfigure[$t$=2640]{\includegraphics[scale=0.5]{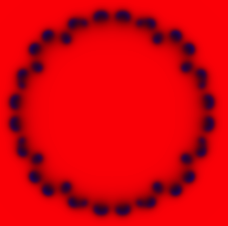}} %
%\subfigure[$t=$]{\includegraphics[scale=0.5]{figs/k_0_049_F_0_010/000002760.png}} %
\subfigure[$t$=2880]{\includegraphics[scale=0.5]{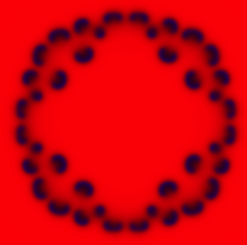}} %
\subfigure[$t$=3010]{\includegraphics[scale=0.5]{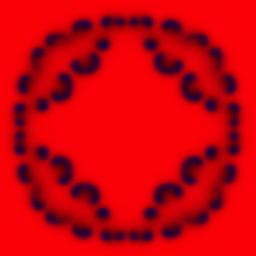}} %
\subfigure[$t$=3330]{\includegraphics[scale=0.5]{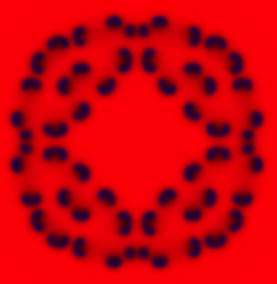}} %
%\subfigure[$t=$]{\includegraphics[scale=0.5]{figs/k_0_049_F_0_010/000004000.png}} %
\subfigure[$t$=6560]{\includegraphics[scale=0.5]{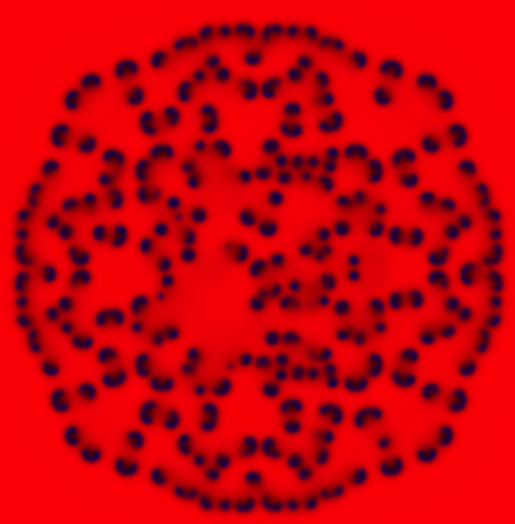}} %
\caption{Snapshots of the evolution of the medium governed by $(k, F)=(0.049, 0.010)$, $t$ is iteration at which the snapshot was recorded. Scale 0.5 of original size.}
\label{k_0_049_F_0_010}
\end{center}
\end{figure}

The pair  $(k, F)=(0.049, 0.010)$ produces most expressive concentration profile, i.e. most complex patterns, as measured by $H$, with least space occupied. 
The medium exhibits a `swarm' of localised travelling patters, soliton-like wave-fragments.  These wave-fragments self-replicate and deflect when collide one with another.  Initial 
perturbation leads to the formation of a `classical' circular wave-front  (Fig.~\ref{k_0_049_F_0_010}a). 
The wave-front loses stability and splits into two centrifugal wave-fragments after 500 iterations (Fig.~\ref{k_0_049_F_0_010}b).
Each of these four fragments splits into two wave-fragments which propagate away from the grid centre and sideways (Fig.~\ref{k_0_049_F_0_010}c). 
The wave-fragments originated from adjacent ends of the parent wave-fragments collide. They deflect in the result of this collision and align their velocity vectors in the centrifugal direction. 
The wave-fragments multiply repeatedly (Fig.~\ref{k_0_049_F_0_010}d)  yet all of them stay on the  expanding circle, forming the beads-like structure (Fig.~\ref{k_0_049_F_0_010}e). 
The order of the wave-fragments breaks up  after $2.5 \cdot 10^3$ iterations (Fig.~\ref{k_0_049_F_0_010}f). The four centripetal wave-fragments emerge (Fig.~\ref{k_0_049_F_0_010}g). 
The centripetal wave-fragments divide (Fig.~\ref{k_0_049_F_0_010}h): their scrolling edges become detached and get transformed into 
wave-fragments propagating centrifugally (Fig.~\ref{k_0_049_F_0_010}i). Eventually the area inside the propagating beads of wave-fragments becomes populated with wave-fragments 
that collide with other wave-fragment,  change their velocity vectors in the result of collisions, split and produce new wave-fragments  (Fig.~\ref{k_0_049_F_0_010}j).

\section{Discussion}

We found that the generative complexity of the Gray-Scott reaction-diffusion medium is due to interacting waves and localised  wave-fragments. 
The Gray-Scott media generating most complex patterns of concentration profiles exhibit wave-fragments and travelling localisations similar to the dissipative solitons. 
These localisations are typically either formed due to circular wave-front gets unstable and breaks up, scrolling of the wave-fragments' ends and formation of 
new wave-fragments. Such dynamics is clearly visible in the media with highest Shannon  entropy and Lempel-Ziv complexity but less apparent in the media with highest 
expressivity, where soliton-like wave-fragments emerge quickly at the first stage of the simulations. The rules with highest values of $H$, $S$ and $E$ 
roughly correspond to Pearson-Munafo class alpha ($\alpha$) (Fig.~\ref{munafoonshannon}), with dynamics described in \cite{munafo2016} as  composed of 
wavelets (aka wave-fragments) and recursively multiplying spirals which annihilate on colliding with each others. Sometimes the behaviour  of the medium is interpreted
as `chaotic'~\cite{nishiura2001spatio, wang2007spatiotemporal} due to irregular deflections of the travelling localisations. 
Rules with highest $LZ$ values roughly correspond to the classe gamma ($\gamma$) --- worm-like branching structures, and epsilon ($\epsilon$) and zeta ($\zeta$) (Fig.~\ref{munafoonshannon}) --- 
unstable travelling multiplying spots  (similar to dissipative solitons)~\cite{munafo2016}.

Our findings on complexity of Gray-Scott media are in agreement with results of our previous studies  on the key role of waves and 
travelling localisations in defining the complexity of spatially-extended non-linear media. Thus, in~\cite{adamatzky2010generative} we constructed a generative 
morphological complexity hierarchy of elementary cellular automata (CA):  one-dimensional CA with three-cell neighbourhoods and binary cells states. 
 Rules with higher generative morphological complexity are Rule 30 and Rule 45~\cite{martinez2013note}. 
 The rules exhibit varieties of travelling localisations, gliders. The gliders collide one with another 
and produce other travelling  localisations in the results of their collisions. Generators of the localisations --- glider guns --- are also observed in the 
space-time configurations generated by Rules 30 and 45~\cite{martinez2009complex, martinez2010make, martinez2012complex}; they are analogs of wave-fragments which produce other wave-fragments. 
Also, in the automaton models of two-species populations~\cite{adamatzky2010minimal} we found that the  basic types of inter-species interactions can be arranged in the following descending 
hierarchy of complexity: mutualism, parasitism,  competition, amensalisms, commensalism. Most complex inter-species interactions show travelling localisations and 
wave-fragments.  In~\cite{adamatzky2012phenomenology} we studied a two-dimensional excitable CA: a resting cell excites if number of excited neighbours lies in a certain
 interval (excitation interval); a refractory cell returns to a resting state only if the number of excited neighbours belong to recovery interval. The model is an excitable
  cellular automaton abstraction of  a spatially extended   semi-memristive medium where a cell's resting state symbolises low-resistance and refractory state high-resistance. 
  We constructed hierarchies of morphological diversity and generative diversity, and found that automata from classes with highest values of complexity quasi-chaotically 
  respond to spatially extended random excitations, develop disordered domains of refractory states filled with breathing cores of localised  excitations and combinations of travelling wave-fronts, wave-fragments and travelling localisations. The automata exhibiting travelling localisations show highest degrees of expressivity. 
  
  In summary, the Gray-Scott media exhibiting waves and, particularly, travelling localisations, or soliton-like wave-fragments, are champions of complexity.

 \section{Supplementary material}
 \label{videos}

 Videos of Gray-Scott model, $768 \times 768$ node grids, frames are saved every 10th iteration, playback is 30 frames per second. 
 Concentrations of $U$ and $V$ in each node $x$ are converted to RGB colour of the corresponding pixel $x$
as $(R, G, B)=(u_x \cdot 255, 0, v_x \cdot 255)$. 
 
 \begin{itemize}
 \item $k=0.048$, $F=0.014$: \url{https://drive.google.com/open?id=0BzPSgPF_2eyUYlFMV3RsbUIxaW8}
\item $k=0.046$, $F=0.011$: \url{https://drive.google.com/open?id=0BzPSgPF_2eyUU2Nsck5PSE5mLW8}
\item $k=0.045$, $F=0.010$: \url{https://drive.google.com/open?id=0BzPSgPF_2eyUdjNmX1l2YV9oVG8}
\item $k=0.048$, $F=0.013$: \url{https://drive.google.com/open?id=0BzPSgPF_2eyUSmRlVTd0R2tSaHc}
\item $k=0.047$, $F=0.013$: \url{https://drive.google.com/open?id=0BzPSgPF_2eyUYmU4b2wtak1sakk}
\item $k=0.049$, $F=0.015$: \url{https://drive.google.com/open?id=0BzPSgPF_2eyUQzdqT29HOVc3bXM}
\item $k=0.062$, $F=0.036$: \url{https://drive.google.com/open?id=0BzPSgPF_2eyUY1BqZFR2UFc3OU0}
\item $k=0.060$, $F=0.027$: \url{https://drive.google.com/open?id=0BzPSgPF_2eyUNll1ajdhYWRJTGc}
\item $k=0.060$, $F=0.026$: \url{https://drive.google.com/open?id=0BzPSgPF_2eyURndKVFVqclh0YmM}
\item $k=0.056$, $F=0.027$: \url{https://drive.google.com/open?id=0BzPSgPF_2eyUcmVPUkRreDIyQlk}
\item $k=0.055$, $F=0.023$: \url{https://drive.google.com/open?id=0BzPSgPF_2eyUVWRoTWFESklDczg}
\item $k=0.058$, $F=0.023$:\url{https://drive.google.com/open?id=0BzPSgPF_2eyUMy1xQTFDTFNiMGc}
\item $k=0.049$, $F=0.010$: \url{https://drive.google.com/open?id=0BzPSgPF_2eyUVGV0VTJ4eWNacFU}
\item $k=0.053$, $F=0.017$: \url{https://drive.google.com/open?id=0BzPSgPF_2eyUTDUzeFo0S1RON3M}
\item $k=0.050$, $F=0.010$: \url{https://drive.google.com/open?id=0BzPSgPF_2eyUTGczelhhRW82SEk}
\end{itemize}

\bibliographystyle{plain}
\bibliography{grayscottbib}

\end{document}